\documentclass[onecolumn,amsmath,amssymb,10pt,aps]{revtex4}
  
\usepackage[utf8]{inputenc}
\usepackage[T1]{fontenc}

\usepackage{amsmath}
\usepackage{amsfonts}
\usepackage{graphicx}
\usepackage{yfonts}
\usepackage{color}
\usepackage[normalem]{ulem}
\usepackage{amsthm}
\usepackage{bm}
\usepackage{bbm}
\usepackage{mathtools}
\usepackage{array}
\usepackage{placeins}
\usepackage{enumitem}

\newcommand{\der}{\,\mathrm{d}}

\newcommand\bigexists{\mbox{\Large $\mathsurround=1pt\exists$}}
\def\<{\langle}
\def\>{\rangle}

\newcommand{\Tr}{\mathrm{Tr}}
\def\oper{{\mathchoice{\rm 1\mskip-4mu l}{\rm 1\mskip-4mu l}
{\rm 1\mskip-4.5mu l}{\rm 1\mskip-5mu l}}}
\DeclareMathAlphabet\mathbfcal{OMS}{cmsy}{b}{n}

\mathchardef\mhyphen="2D 

\begin{document}

\title{Engineering classical capacity of generalized Pauli channels with admissible memory kernels}

\author{Katarzyna Siudzi{\'n}ska, Arpan Das, and Anindita Bera}
\affiliation{Institute of Physics, Faculty of Physics, Astronomy and Informatics \\  Nicolaus Copernicus University, ul. Grudzi\k{a}dzka 5/7, 87--100 Toru{\'n}, Poland}

\begin{abstract}
We analyze the classical capacity of the generalized Pauli channels generated via the memory kernel master equations. For suitable engineering of the kernel parameters, the evolution with non-local noise effects can produce dynamical maps with higher capacity than the purely Markovian evolution. We provide instructive examples for qubit and qutrit evolution. Interestingly, similar behavior is not observed when analyzing the time-local master equations.
\end{abstract}

\flushbottom

\maketitle

\thispagestyle{empty}

\section{Introduction}

In quantum information processing, it is crucial to understand how to transmit, manipulate, and preserve quantum information sent through a noisy quantum channel \cite{Nielsen,Bruss}. Due to scientific and technological advancements, logic gates and other electronic devices are approaching atomic scales. Therefore, it becomes increasingly hard to reliably transfer information. This can be remedied if one minimizes detrimental effects of noise through error correction, error mitigation, or error suppression techniques \cite{Review,QEC}. 

However, removing errors is only one way to deal with undesirable effects of environmental noise on quantum systems. Another approach to the problem is, instead of reducing the noise, using it to one's advantage. This perception on the role of environmental noise was started by the observation that dissipation can be used to enhance quantum information processing \cite{Verstraete}. This way, dissipation became a quantum resource exploited to manipulate quantum systems and engineer specific properties of quantum channels \cite{Zanardi,Marshall,Gillard}. In particular, memory effects caused by environmental noise were used for performing quantum information processing tasks, like improving channel fidelity or preserving quantum entanglement \cite{zanardi17}. A decrease of error accumulation was achieved for the dissipative Markovian processes and their generalizations \cite{Shabani,semi-2}, where adding noises to the Markovian evolution slows down the rate at which the state of the system approaches the steady state.

The goal of this paper is showing how to engineer quantum noise to improve the channel capacity, which is a very important measure considered in quantum computation and quantum information theory. With the channel capacity, one determines the amount of information transmitted coherently through a quantum channel. However, in contrast to the classical channels with the unique (Shannon) capacity, the concept of quantum channel capacity is more complex, giving rise to a whole range of informational characteristics. If quantum information is transferred through a noisy channel, then one considers the quantum capacity, whose lower and upper estimations were found by Lloyd \cite{Lloyd}, Shor \cite{Shor}, and Devetak \cite{Devetak2}. In quantum cryptography, communication tasks often require the use of private classical capacity \cite{Devetak2}. Also, quantum correlations are essential for the entanglement-assisted capacity \cite{Bennett}, which is the highest rate of classical information transition. The problem of simultaneously transferring classical and quantum information was investigated by Devetak and Shor \cite{ShorDevetak}. More information about channel capacities is available in review works, see e.g. \cite{Gyongyosi,Smith}.

The capacity that directly generalizes the notion of Shannon capacity for classical channels to the quantum scenario is the classical capacity \cite{Holevo,sw}. In this case, classical information is sent through a quantum channel using separable input states and joint measurements on the outputs. Recently, there has been significant interest in calculating classical capacity of quantum channels. Rehman et al. used the majorization procedure to provide lower and upper estimations on the Holevo capacity of the Weyl channels \cite{WCHC,WCHC2}. Amosov calculated the classical capacity for deformations of classical-quantum Weyl channels \cite{Amosov4} and channels generated by irreducible projective unitary representations of finite groups \cite{Amosov5}.

In this paper, we analyze a time evolution of the classical capacity for the generalized Pauli channels \cite{Ruskai,mub_final}. In particular, we compare the capacity for the dynamical maps governed by the memory kernel
\begin{equation}
K(t)=\mathcal{L}\delta(t)+\mathbb{K}(t)
\end{equation}
and with the Markovian generator $\mathcal{L}$ alone. In the above formula, $\mathbb{K}(t)$ is the part of the kernel that does not involve the local part with the Dirac delta function $\delta(t)$. With a proper choice of parameters, we propose a number of cases where the classical capacity of the map generated by $K(t)$ is better than that of the Markovian semigroup $\Lambda^{\rm M}(t)=e^{t\mathcal{L}}$. Hence, it is shown that non-local memory effects can be used effectively to decrease the error rate of a quantum channel. We also present a class of quantum evolution where the generator $\mathcal{L}(t)$ is time-local. This implies that improving the channel capacity is possible not only for the Markovian semigroup but for a general Markovian dynamics.

\section{Generalized Pauli channels}

An important class of quantum channels consists of mixed unitary channels, where a unitary evolution is disrupted by classical errors \cite{TQI,Alicki}. The channel noise can be corrected with the classical information obtained by measuring the environment \cite{Gregoratti}. For qubit systems, one considers the Pauli channel \cite{King,Landau}
\begin{equation}\label{Pauli}
\Lambda[\rho]=\sum_{\alpha=0}^3 p_\alpha \sigma_\alpha \rho \sigma_\alpha,
\end{equation}
where $p_\alpha$ is a probability distribution and $\sigma_0=\mathbb{I}_2,\sigma_1,\sigma_2,\sigma_3$ are the Pauli matrices. As the Kraus representation of a quantum map is not unique, it is often more convenient to work with its spectrum. One finds the eigenvalues of the Pauli channel through its eigenvalue equations
\begin{equation}
\Lambda[\sigma_\alpha]=\lambda_\alpha\sigma_\alpha,\qquad \lambda_0=1.
\end{equation}
An important property of $\sigma_\alpha$, $\alpha=1,2,3$, is that their eigenvectors $\{\psi_0^{(\alpha)},\psi_1^{(\alpha)}\}$ form three mutually unbiased bases (MUBs). Recall that two orthonormal bases are mutually unbiased if and only if
\begin{equation}
\big|\big\<\psi_k^{(\alpha)}\big|\psi_l^{(\beta)}\big\>\big|^2=
\frac 1d
\end{equation}
for $\alpha\neq\beta$ and $k,l=0,\ldots,d-1$, where $d$ is the dimension of the underlying Hilbert space ($d=2$ for qubits).

The Pauli channels can be generalized in multiple ways \cite{Petz,Filip2,Gell-Mann,MUM_GPC}, but only one generalization ensures that the MUB property of its eigenvectors carries over to $d>2$. Consider the $d$-dimensional Hilbert space $\mathcal{H}$ that admits the maximal number of $d+1$ mutually unbiased bases \cite{MAX}. Using the rank-1 projectors $P_k^{(\alpha)}:=|\psi_k^{(\alpha)}\>\<\psi_k^{(\alpha)}|$, one defines $d^2-1$ unitary operators
\begin{equation}\label{U}
U_\alpha^k=\sum_{l=0}^{d-1}\omega^{kl}P_l^{(\alpha)},\qquad \omega:=e^{2\pi i/d}.
\end{equation}
The generalized Pauli channel is constructed as follows \cite{Ruskai,mub_final},
\begin{equation}\label{GPC}
\Lambda[\rho]=p_0\rho+\frac{1}{d-1}\sum_{\alpha=1}^{d+1}p_\alpha\sum_{k=1}^{d-1}
U_\alpha^k\rho U_\alpha^{k\dagger},
\end{equation}
where the Pauli channel in Eq. (\ref{Pauli}) is reproduced after setting $d=2$. The eigenvalues $\lambda_\alpha$ of $\Lambda$ are real and $(d-1)$-times degenerated. They satisfy the eigenvalue equations
\begin{equation}\label{GPC_eigenvalue_eq}
\Lambda[U_\alpha^k]=\lambda_\alpha U_\alpha^k,\qquad k=1,\ldots,d-1,
\end{equation}
and $\Lambda[\mathbb{I}_d]=\mathbb{I}_d$. In terms of the probability distribution $p_\alpha$,
\begin{equation}\label{GPC_eigenvalues}
\lambda_\alpha=\frac{1}{d-1}\left[d(p_0+p_\alpha)-1\right],
\end{equation}
whereas the inverse relation reads
\begin{equation}\label{CCC}
\begin{split}
p_0&=\frac{1}{d^2}\left(1+(d-1)\sum_{\alpha=1}^{d+1}\lambda_\alpha\right),\\
p_\alpha&=\frac{d-1}{d^2}\left(1+d\lambda_\alpha-\sum_{\beta=1}^{d+1} \lambda_\beta\right).
\end{split}
\end{equation}
The complete positivity of the generalized Pauli channel is fully controlled by its eigenvalues. Indeed, $\Lambda$ is completely positive if and only if $\lambda_\alpha$ satisfy the generalized Fujiwara-Algoet conditions \cite{Fujiwara,Ruskai,Zyczkowski}
\begin{equation}\label{Fuji-d}
-\frac{1}{d-1}\leq\sum_{\beta=1}^{d+1}\lambda_\beta\leq 1+d\min_{\beta>0}\lambda_\beta.
\end{equation}

\section{Classical capacity of generalized Pauli channels}

In the classical theory of information, there exists a unique measure for the amount of information reliably transmitted through a noisy channel. This measure is known as the Shannon capacity, and it is a maximization of the mutual information between its input and output states over all random variable probability distributions \cite{Shannon}. In quantum information theory, however, information can be transmitted in a number of ways. Therefore, there exist many types of channel capacities, such as the quantum capacity \cite{Lloyd,Shor,Devetak2}, private classical capacity \cite{Devetak2}, or entanglement-assisted capacity \cite{Bennett}. A direct analogue of the Shannon capacity in the quantum scenario is the Holevo capacity. It determines the maximal amount of classical information that can be reliably transferred, provided that the input state is separable and the output state is measured via joint measurements \cite{KingRemarks,Gyongyosi}. The Holevo capacity is defined as the maximal value of the entropic expression \cite{Holevo,sw}
\begin{equation}
\chi(\Lambda)=\max_{\{p_k,\rho_k\}}\left[S\left(\sum_kp_k\Lambda[\rho_k]\right)
-\sum_kp_kS(\Lambda[\rho_k])\right],
\end{equation}
where $\Lambda$ is a quantum channel, and $S(\rho):=-\Tr(\rho\ln\rho)$ denotes the von Neumann entropy. Note that the maximum is calculated over the ensembles of separable states $\rho_k$ with the probabilities of occurence $p_k$. The optimal transition rate under infinitely many uses of a channel is given by the classical capacity
\begin{equation}
C(\Lambda)=\lim_{n\to\infty}\frac 1n \chi(\Lambda^{\otimes n}).
\end{equation}
In general, $C(\Lambda)\geq\chi(\Lambda)$. However, for a weakly additive Holevo capacity ($\chi(\Lambda\otimes\Lambda)=2\chi(\Lambda)$), one has $C(\Lambda)=\chi(\Lambda)$ \cite{sw}.

In ref. \cite{Holevo_capacity}, exact values of the classical capacity were found for certain families of the generalized Pauli channels.
Namely, if all $\lambda_\alpha\leq 0$, and moreover $\lambda_1=\ldots=\lambda_d\equiv\lambda_{\max}$, $\lambda_{d+1}=\lambda_{\min}$, then
\begin{equation}\label{11}
C(\Lambda)=\frac{1+(d-1)\lambda_{\min}}{d}\ln[1+(d-1)\lambda_{\min}]
+(d-1)\frac{1-\lambda_{\min}}{d}\ln(1-\lambda_{\min}).
\end{equation}
On the contrary, if all $\lambda_\alpha\geq 0$ and also $\lambda_1=\lambda_{\max}$, $\lambda_2=\ldots=\lambda_{d+1}\equiv\lambda_{\min}$, then
\begin{equation}\label{22}
C(\Lambda)=\frac{1+(d-1)\lambda_{\max}}{d}\ln[1+(d-1)\lambda_{\max}]
+(d-1)\frac{1-\lambda_{\max}}{d}\ln(1-\lambda_{\max}).
\end{equation}
In addition, if all of the eigenvalues are equal to one another, so that $\lambda_1=\ldots=\lambda_{d+1}\equiv\lambda$, then one recovers the capacity of the depolarizing channel \cite{King2}. For any other combination of eigenvalues, one finds only the lower bound of the capacity \cite{Holevo_capacity},
\begin{equation}\label{33}
C_{\rm low}(\Lambda)=\max_{\alpha>0}c_\alpha,\qquad
c_\alpha=\frac{1+(d-1)\lambda_\alpha}{d}\ln[1+(d-1)\lambda_\alpha]
+\frac{d-1}{d}(1-\lambda_\alpha)\ln(1-\lambda_\alpha).
\end{equation}
In the special case of $d=2$ (the Pauli channels), there above formula gives the exact value of the capacity \cite{WCHC}, so that $C(\Lambda)=C_{\rm low}(\Lambda)$.

\subsection{Generators vs. memory kernels}

The evolution $\rho\longmapsto\rho(t)=\Lambda(t)[\rho]$ of an open quantum system is described by a family of time-parameterized quantum channels $\Lambda(t)$, $t\geq 0$, with the initial condition $\Lambda(0)=\oper$. Such maps can be obtained as solutions to the master equations. In the simplest scenario, the evolution equation $\dot{\Lambda}(t)=\mathcal{L}\Lambda(t)$, where $\mathcal{L}$ is the Gorini-Kossakowski-Sudarshan-Landblad (GKSL) generator \cite{GKS,L}. The solution of such equation is the Markovian semigroup $\Lambda(t)=\exp(t\mathcal{L})$. For the generalized Pauli channels, one has \cite{mub_final}
\begin{equation}
\mathcal{L}=\sum_{\alpha=1}^{d+1}\gamma_\alpha\mathcal{L}_\alpha
\end{equation}
with the decoherence rates $\gamma_\alpha\geq 0$ and
\begin{equation}\label{La}
\mathcal{L}_\alpha[\rho]=\frac 1d \left[\sum_{k=1}^{d-1}U_\alpha^k\rho U_\alpha^{k\dagger}-(d-1)\rho\right].
\end{equation}

Generators that are constant in time are sufficient for open system dynamics with weak coupling to the environment. When this coupling is relatively strong, however, it becomes essential to consider the master equations that take non-Markovian memory effects into account. One generalization of the semigroup master equation is $\dot{\Lambda}(t)=\mathcal{L}(t)\Lambda(t)$, where the constant generator is replaced with the time-local generator $\mathcal{L}(t)$. For the generalized Pauli channels, one has
\begin{equation}\label{MS}
\mathcal{L}(t)=\sum_{\alpha=1}^{d+1}\gamma_\alpha(t)\mathcal{L}_\alpha.
\end{equation}
The condition on the decoherence rates is relaxed, as they no longer have to be positive for the dynamics to be legitimate. This time, $\gamma_\alpha\geq 0$ is the necessary and sufficient condition for the corresponding (invertible) $\Lambda(t)$ to be Markovian in terms of divisibility \cite{RHP,Wolf}. A dynamical map is CP-divisible if and only if it is decomposable into $\Lambda(t)=V(t,s)\Lambda(s)$ for any $t\geq s\geq 0$. The propagator $V(t,s)$ is then a completely positive, trace-preserving map.

By solving the evolution equation with the time-local generator, we find that the eigenvalues of the associated dynamical map read \cite{mub_final}
\begin{equation}
\lambda_\alpha(t)=\exp[\Gamma_\alpha(t)-\Gamma_0(t)],
\end{equation}
where $\Gamma_\alpha(t)=\int_0^t\gamma_\alpha(\tau)\der\tau$ for $\alpha=0,\ldots,d+1$ and $\gamma_0(t)=\sum_{\alpha=1}^{d+1}\gamma_\alpha(t)$. Note that the complete positivity conditions from Eq. (\ref{Fuji-d}) reduce to
\begin{equation}
\sum_{\alpha=1}^{d+1} e^{\Gamma_\alpha(t)} \leq e^{\Gamma(t)}+d \min_\beta e^{\Gamma_\beta(t)}.
\end{equation}

Another generalization of the Markovian semigroup master equation is realized using memory kernels. In this approach, the GKSL generator gets replaced with an integral expression. Now, the evolution of the system is governed by the Nakajima-Zwanzig equation \cite{Nakajima,Zwanzig}
\begin{equation}
\label{ker1}
\dot{\Lambda}(t)=\int_0^t K(t-\tau) \Lambda(\tau)\der\tau,
\end{equation}
where $K(t)$ is the memory kernel. Observe that this is an integro-differential equation, and therefore the evolved state $\rho(t)$ depends on every earlier state $\rho(\tau)$, $\tau<t$. The memory kernel that corresponds to the generalized Pauli channels has a relatively simple form,
\begin{equation}
\label{ker2}
K(t)=\sum_{\alpha=1}^{d+1} k_\alpha(t) \mathcal{L}_\alpha.
\end{equation}
Note that $K(t)$ and $\Lambda(t)$ have common eigenvectors,
\begin{equation}
K(t)[U_\alpha^k]=\kappa_\alpha(t) U_\alpha^k,\qquad K(t)[\mathbb{I}]=0,
\end{equation}
where
\begin{equation}
\kappa_\alpha(t)=k_\alpha(t)-k_0(t)
\end{equation}
with $k_0(t)=\sum_\beta^{d+1} k_\beta(t)$ are the eigenvalues of the kernel.
Hence, one can rewrite the Nakajima-Zwanzig equation as
\begin{equation}
\dot{\lambda}_\alpha(t)=\int_0^t \kappa_\alpha(t-\tau) \lambda_\alpha(\tau) \der\tau.
\end{equation}
In the Laplace transform domain, the solution reads
\begin{equation}
\tilde{\lambda}_\alpha(s)=\frac{1}{s-\tilde{\kappa}_\alpha(s)},
\end{equation}
where $\tilde{f}(s)=\int_0^\infty f(t)e^{-st}dt$ is the Laplace transform of the function $f(t)$.

The necessary and sufficient conditions for legitimate memory kernels were provided in ref. \cite{memory}. First, one parameterizes the eigenvalues $\lambda_\alpha(t)$ of the dynamical map by a real function $\ell_\alpha(t)$ in such a way that
\begin{equation}
\lambda_\alpha(t)=1-\int_0^t \ell_\alpha(\tau)\der\tau.
\end{equation}
Now, the associated kernel is legitimate if and only if its eigenvalues
\begin{equation}
\tilde{\kappa}_\alpha(s)=-\frac{s \tilde{\ell}_\alpha(s)}{1-\tilde{\ell}_\alpha(s)},
\end{equation}
where $\ell_\alpha(t)$ satisfy the additional conditions
\begin{align}
\int_0^t \ell_\alpha(\tau)\der\tau&\geq 0,\\
d \int_0^t \ell_\alpha(\tau)\der\tau&\leq
\sum_{\beta=1}^{d+1} \int_0^t \ell_\beta(\tau) \der\tau \leq \frac{d^2}{d-1},
\end{align}
for $\alpha=1,2,\ldots,d+1$.

\section{Engineering capacity through kernel manipulations}

In this section, we analyze how the classical capacity of the generalized Pauli channels changes in time for the evolution generated by Eq. (\ref{ker1}) with the memory kernel
\begin{equation}\label{kernel}
K(t)=\delta(t)\mathcal{L}+\mathbb{K}(t).
\end{equation}
Notably, in the formula above, $\mathcal{L}$ is a legitimate Markovian semigroup generator from Eq. (\ref{MS}), and $\mathbb{K}(t)$ is a legitimate, purely non-local memory kernel (i.e., it does not involve the Dirac delta function $\delta(t)$). It is shown that, by adding a non-local part $\mathbb{K}(t)$, one can improve the classical capacity of the associated dynamical map $\Lambda(t)$. 

The addition of purely local and non-local kernels was already considered in refs. \cite{zanardi17,fidelity}, where it was proved that the channel fidelity can temporarily increased by an appropriate engineering of kernel parameters. 
In what follows, we consider three types of dynamical maps: the Markovian semigroup $\Lambda^{\rm M}(t)=e^{t\mathcal{L}}$, the non-Markovian noise $\Lambda^{\rm N}(t)$ that solves $\dot{\Lambda}^{\rm N}(t)=\int_0^t \mathbb{K}(t-\tau) \Lambda^{\rm N}(\tau)\der\tau$, and finally the map $\Lambda(t)$ that satisfies the Nakajima-Zwanzig equation with $K(t)=\delta(t)\mathcal{L}+\mathbb{K}(t)$. The eigenvalues of the corresponding maps are denoted by $\lambda^{\rm M}_\alpha(t)$, $\lambda^{\rm N}_\alpha(t)$, and $\lambda_\alpha(t)$, respectively. Interestingly, there is no simple relation between the map eigenvalues, as in the Laplace transform domain
\begin{equation}
\widetilde{\lambda}_\alpha(s)=\frac{\widetilde{\lambda}^{\rm M}_\alpha(s)\widetilde{\lambda}^{\rm N}_\alpha(s)}
{\widetilde{\lambda}^{\rm M}_\alpha(s)+\widetilde{\lambda}^{\rm N}_\alpha(s)
-s\widetilde{\lambda}^{\rm M}_\alpha(s)\widetilde{\lambda}^{\rm N}_\alpha(s)}.
\end{equation}
In the following examples, the map that describes the noise part is always non-invertible and not kernel non-decreasing; that is,
\begin{equation}
\bigexists {0\leq \tau\leq t}:\quad\mathrm{ker}\Lambda^{\rm N}(\tau)\nsubseteq
\mathrm{ker}\Lambda^{\rm N}(t).
\end{equation}
In other words, there exists at least one eigenvalue $\lambda_\alpha^{\rm N}(t)$ that reaches zero at some finite time $t_\ast$ but does not remain zero for some $t>t_\ast$. Such dynamical maps are indivisible, and hence the corresponding evolution is non-Markovian \cite{div_inf_flow}.

\subsection{Constant kernel}

First, consider the qubit evolutions ($d=2$) provided by the isotropic Markovian generator
\begin{equation}
\mathcal{L}=\frac{\gamma}{2}\sum_{\alpha=1}^3\mathcal{L}_\alpha
\end{equation}
with a positive decoherence rate $\gamma$ and the memory kernel $\mathbb{K}(t)$ with constant eigenvalues
\begin{equation}
\kappa^{\rm N}_1(t)=\kappa^{\rm N}_2(t)=-\omega^2,\qquad 
\kappa^{\rm N}_3(t)=0,
\end{equation}
where $\omega>0$.
The corresponding solutions read
\begin{equation}
\lambda^{\rm M}_1(t)=
\lambda^{\rm M}_2(t)=\lambda^{\rm M}_3(t)=e^{-\gamma t},
\end{equation}
and
\begin{equation}
\lambda^{\rm N}_1(t)=\lambda^{\rm N}_2(t)=\cos\omega t,\qquad 
\lambda^{\rm N}_3(t)=1,
\end{equation}
respectively. Observe that the dynamical maps characterized via $\lambda^{\rm M}_\alpha(t)$ and $\lambda^{\rm N}_\alpha(t)$ are always legitimate. 

\begin{figure}[htb!]
 \includegraphics[width=0.6\textwidth]{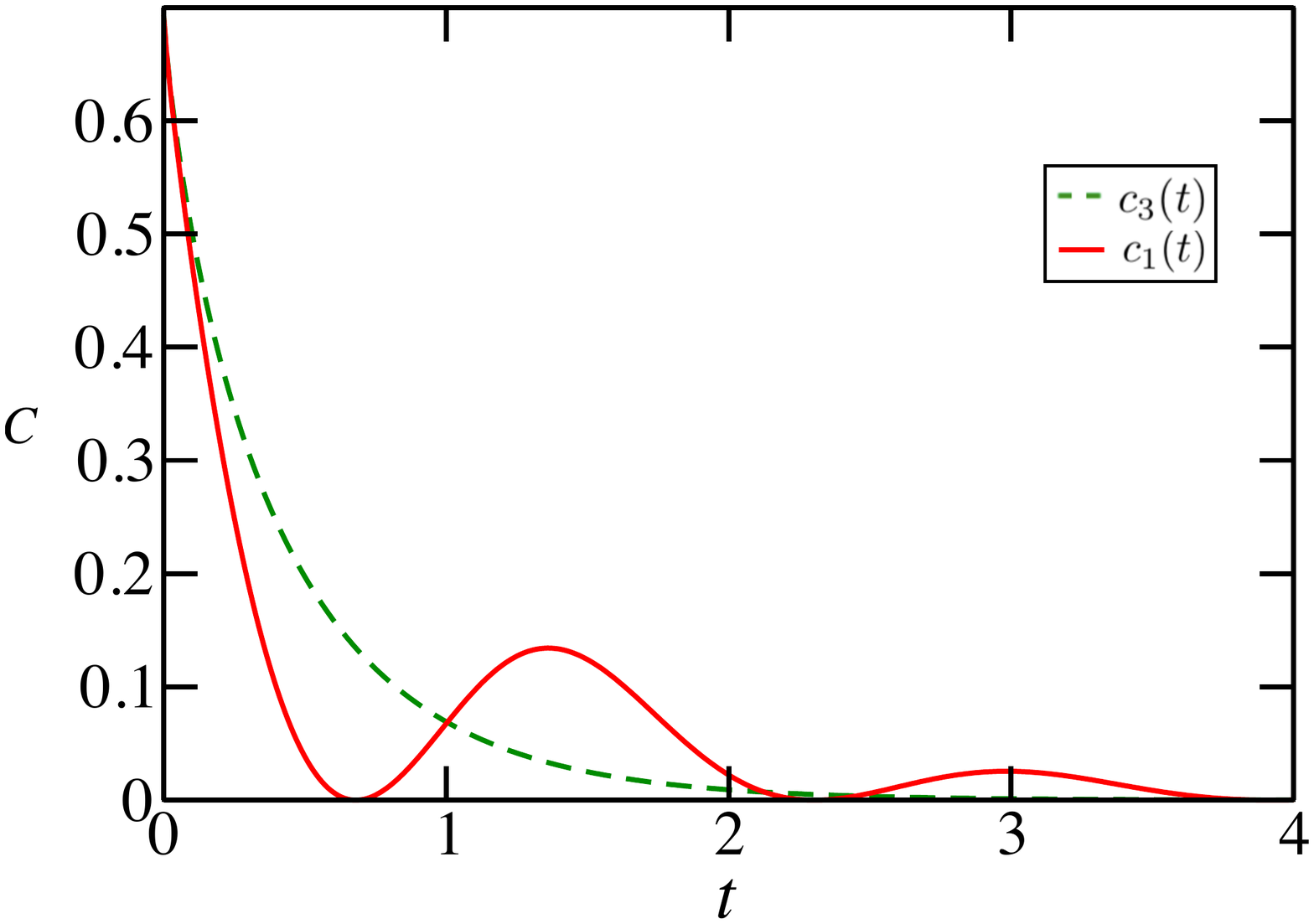}
\caption{
Illustration of an increase in classical capacity for the qubit evolution with $\gamma=1/\mathrm{s}$ and $\omega=2/\mathrm{s}$.
}
\label{kernel1}
\end{figure}

The Pauli dynamical map generated by $K(t)=\delta(t)\mathcal{L}+\mathbb{K}(t)$ is characterized by the following eigenvalues,
\begin{equation}\label{L1}
\lambda_1(t)=\lambda_2(t)=\frac{2\omega}{P}e^{-\gamma t/2}
\cos\left(\frac{Pt}{2}+\arctan\frac{\gamma}{P}\right),\qquad
\lambda_3(t)=e^{-\gamma t},
\end{equation}
where $P=\sqrt{4\omega^2-\gamma^2}$. The eigenvalues $\lambda_1(t)$ and $\lambda_2(t)$ oscillate if and only if $\gamma<2\omega$, which is also a necessary condition for complete positivity of $\Lambda(t)$. Additionally, for $\Lambda(t)$ to describe a legitimate evolution, it is sufficient that
\begin{equation}
\cosh\frac{\gamma t_\ast}{2}\geq\frac{2\omega}{P},
\end{equation}
where
\begin{equation}\label{tast}
t_\ast=\frac{2}{P}\left(\pi-\arctan\frac{\gamma}{P}\right)
\end{equation}
is the time corresponding to the first local minimum of the cosine function. This is a direct consequence of the Fujiwara-Algoet conditions from Eq. (\ref{Fuji-d}).
Hence, a combination of two legitimate memory kernels does not necessary yield a physical dynamics. Now, using Eq. (\ref{33}), we calculate the classical capacity of $\Lambda(t)$,
\begin{equation}\label{n1}
C[\Lambda(t)]=\max\Big\{c_1(t),c_3(t)\Big\},
\end{equation}
where $c_1(t)=c_2(t)$ and $c_3(t)=C[\Lambda^{\rm M}(t)]$. Therefore, whenever $c_1(t)>c_3(t)$, one observes an increase of capacity for the system with additional noises. An exemplary choice of parameters is shown in Fig. \ref{kernel1}.

\subsection{Exponential decay}

Let us take the Markovian semigroup generated by
\begin{equation}
\mathcal{L}=\frac{\gamma}{d}\sum_{\alpha=1}^{d+1}\mathcal{L}_\alpha
\end{equation}
and the exponentially decaying memory kernel $\mathbb{K}(t)$, similar to the one analyzed in \cite{zanardi17,fidelity}, with
\begin{equation}\label{ker}
\kappa^{\rm N}_{\alpha}(t)=-\omega^2e^{-Zt};\qquad \kappa^{\rm N}_{\alpha_\ast}(t)=0,\qquad \alpha\neq\alpha_\ast.
\end{equation}
Assume that the constant $\gamma$, $Z$, and $\omega$ are positive.
By solving the master equations, one finds the associated dynamical maps $\Lambda^{\rm M}(t)$ and $\Lambda^{\rm N}(t)$, whose eigenvalues are given by
\begin{equation}
\lambda^{\rm M}_\alpha(t)=e^{-\gamma t}
\end{equation}
and $\lambda^{\rm N}_{\alpha_\ast}(t)=1$,
\begin{equation}\label{L2}
\lambda^{\rm N}_{\alpha}(t)=\frac{2\omega}{P}e^{-Zt/2}
\cos\left(\frac{Pt}{2}-\arctan\frac{Z}{P}\right)
\end{equation}
for $\alpha\neq\alpha_\ast$, where $P=\sqrt{4\omega^2-Z^2}$. Note that for $Z=\gamma$, Eq. (\ref{L2}) is very similar to $\lambda_1(t)$ from Eq. (\ref{L1}) but differs in the sign before arcus tangent. The map $\Lambda^{\rm M}(t)$ is always legitimate, whereas $\Lambda^{\rm N}(t)$ describes a physical dynamics if
\begin{equation}\label{CPC}
e^{Zt_\ast/2}\geq\frac{2(d-1)\omega}{P},
\end{equation}
where
\begin{equation}\label{tast2}
t_\ast=\frac{2}{P}\left(\pi+\arctan\frac{Z}{P}\right)
\end{equation}
corresponds to the first local minimum of the cosine function.

\begin{figure}[htb!]
\includegraphics[width=0.6\textwidth]{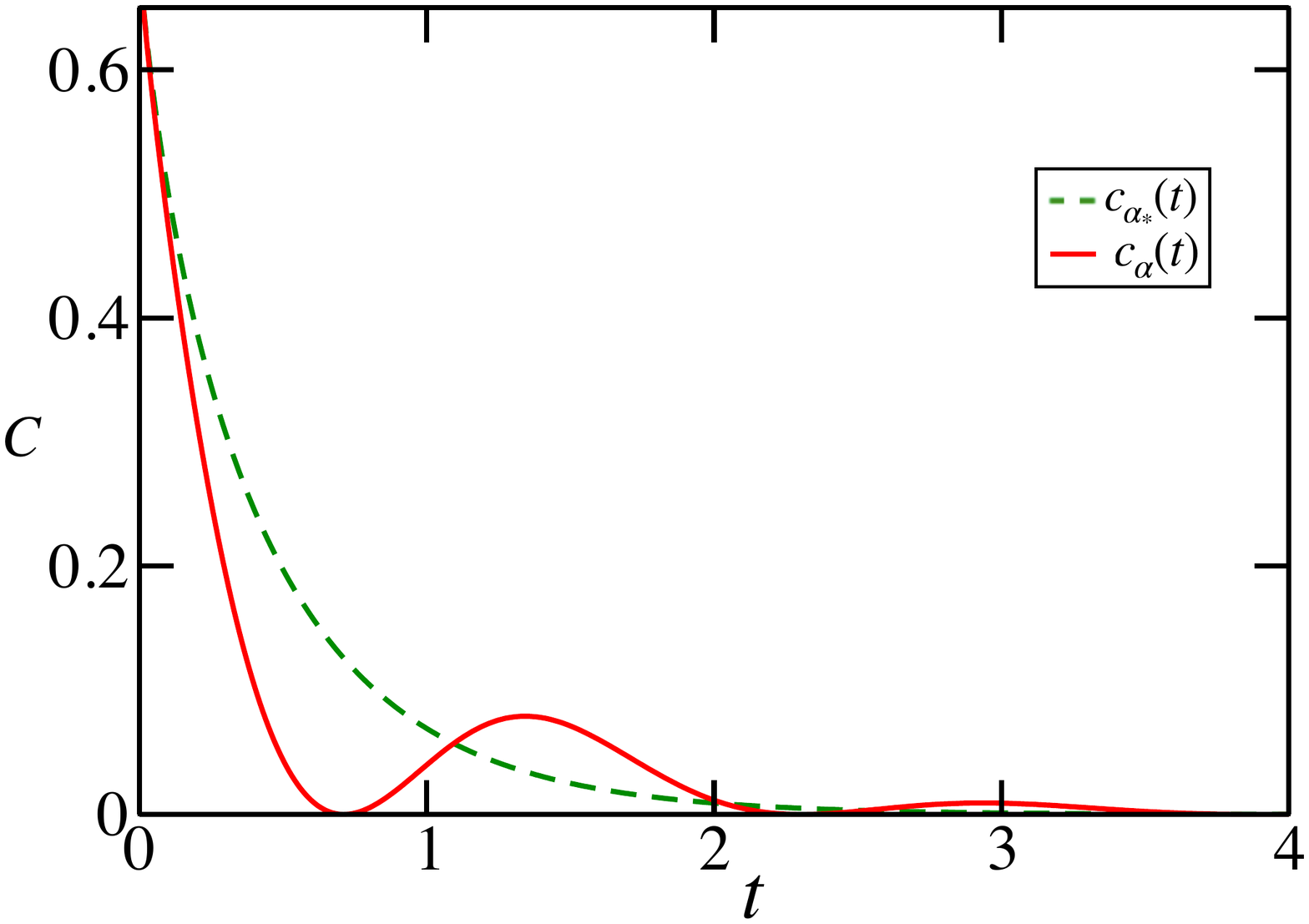}
\caption{
Illustration of an increase in classical capacity for the qubit evolution with  $\gamma=1/\mathrm{s}$, $Z=1/(3\mathrm{s})$, and $\omega=2/\mathrm{s}$.
}
\label{kernel2}
\end{figure}

\begin{figure}[htb!]
\includegraphics[width=0.6\textwidth]{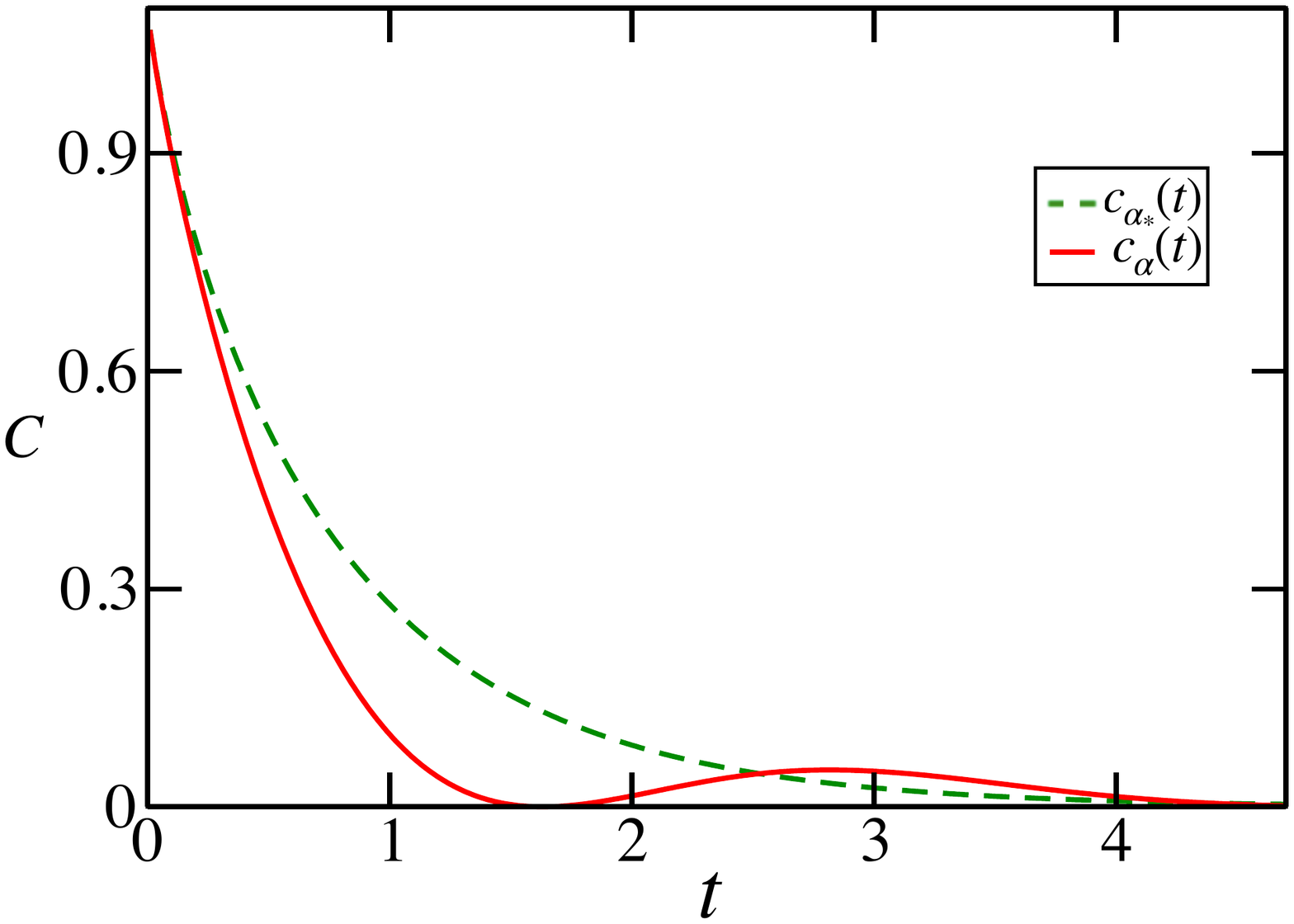}
\caption{
Illustration of an increase in classical capacity for the qutrit evolution with  $\gamma=3/(5\mathrm{s})$, $Z=1/(5\mathrm{s})$, and $\omega=19/(20\mathrm{s})$.
}
\label{kernel3d}
\end{figure}

Now, we analyze the behavior of the dynamical map obtained using $K(t)=\delta(t)\mathcal{L}+\mathbb{K}(t)$. Namely, after adding the non-Markovian noise to the semigroup, the eigenvalue $\lambda_{\alpha_\ast}(t)=e^{-\gamma t}$ remains unchanged. On the other hand,
\begin{equation}\label{LL2}
\lambda_{\alpha}(t)=\frac{2\omega}{R}e^{-(\gamma+Z)t/2}
\cos\left(\frac{Rt}{2}+\arctan\frac{\gamma-Z}{R}\right),
\end{equation}
for $\alpha\neq\alpha_\ast$, where $R=\sqrt{4\omega^2-(\gamma-Z)^2}$. Note that  Eq. (\ref{LL2}) is not a simple shift of Eq. (\ref{L2}) by $Z\longmapsto\gamma-Z$, as there are additionally two sign differences.
A sufficient condition for $\Lambda(t)$ to produce a legitimate evolution is
\begin{equation}
\frac{2d(d-1)\omega}{R}\leq e^{Zt_\ast/2}\left[(d-1)e^{-\gamma t_\ast/2}+e^{\gamma t_\ast/2}\right],
\end{equation}
where this time
\begin{equation}
t_\ast=\frac{2}{R}\left(\pi-\arctan\frac{\gamma-Z}{R}\right).
\end{equation}
Assuming that $\Lambda(t)$ describes a qudit evolution, Eq. (\ref{33}) gives the following formula for the lower bound of the classical capacity of $\Lambda(t)$,
\begin{equation}
C[\Lambda(t)]=\max\Big\{c_\alpha(t),c_{\alpha_\ast}(t)\Big\}.
\end{equation}
Observe that $c_{\alpha_\ast}(t)=C[\Lambda^{\rm M}(t)]$, and hence the channel capacity for $\Lambda(t)$ is greater than for the Markovian evolution if $c_\alpha(t)>c_{\alpha_\ast}(t)$. Two examples of appropriate parameter engineering are presented in Fig. \ref{kernel2} for $d=2$ and Fig. \ref{kernel3d} for $d=3$.

\FloatBarrier

\subsection{Beyond the semigroup}

The classical capacity can also be enhanced in a more general case. Let us consider the Markovian evolution is characterized by a dynamical map $\Lambda^{\rm M}(t)$ that is not a semigroup. Instead, it is generated via the time-local generator $\mathcal{L}^{\rm M}(t)$ from Eq. (\ref{MS}) with $\gamma_\alpha^{\rm M}(t)\geq 0$. Now, the most natural way to introduce noises is to add the generator $\mathcal{L}^{\rm N}(t)$ of a non-Markovian evolution, where at least one decoherence rate $\gamma_\alpha^{\rm N}(t)\ngeq 0$. The resulting dynamical map $\Lambda(t)$ is provided via
\begin{equation}
\mathcal{L}(t)=\mathcal{L}^{\rm M}(t)+\mathcal{L}^{\rm N}(t).
\end{equation}
From a physical point of view, one can add two legitimate
generators when the environmental cross-correlations can be ignored \cite{Kolodynski}. Now, the eigenvalues of the generalized Pauli map $\Lambda(t)$ read
\begin{equation}
\lambda_\alpha(t)=\lambda_\alpha^{\rm M}(t)\lambda_\alpha^{\rm N}(t),
\end{equation}
which means that $\Lambda(t)=\Lambda^{\rm M}(t)\Lambda^{\rm N}(t)$ is a composition of two (commutative) generalized Pauli dynamical maps. However, due to $\lambda_\alpha(t)\geq 0$ for any $\Lambda(t)$ that arises from a legitimate time-local generator, $\lambda_\alpha(t)\leq\lambda_\alpha^{\rm M}(t)$, and therefore there can be no increase in the classical capacity.
Therefore, let us instead consider a more general form of the memory kernel $K(t)$. Namely, replace the semigroup generator $\mathcal{L}$ in Eq. (\ref{kernel}) with the memory kernel $\mathfrak{K}(t)$ that describes the same evolution as the time-local generator $\mathcal{L}(t)$. Then, one has
\begin{equation}
K(t)=\mathfrak{K}(t)+\mathbb{K}(t),
\end{equation}
where $\mathfrak{K}(t)$ and $\mathbb{K}(t)$ correspond to a Markovian and non-Markovian dynamics, respectively.

\begin{figure}[htb!]
  \includegraphics[width=0.6\textwidth]{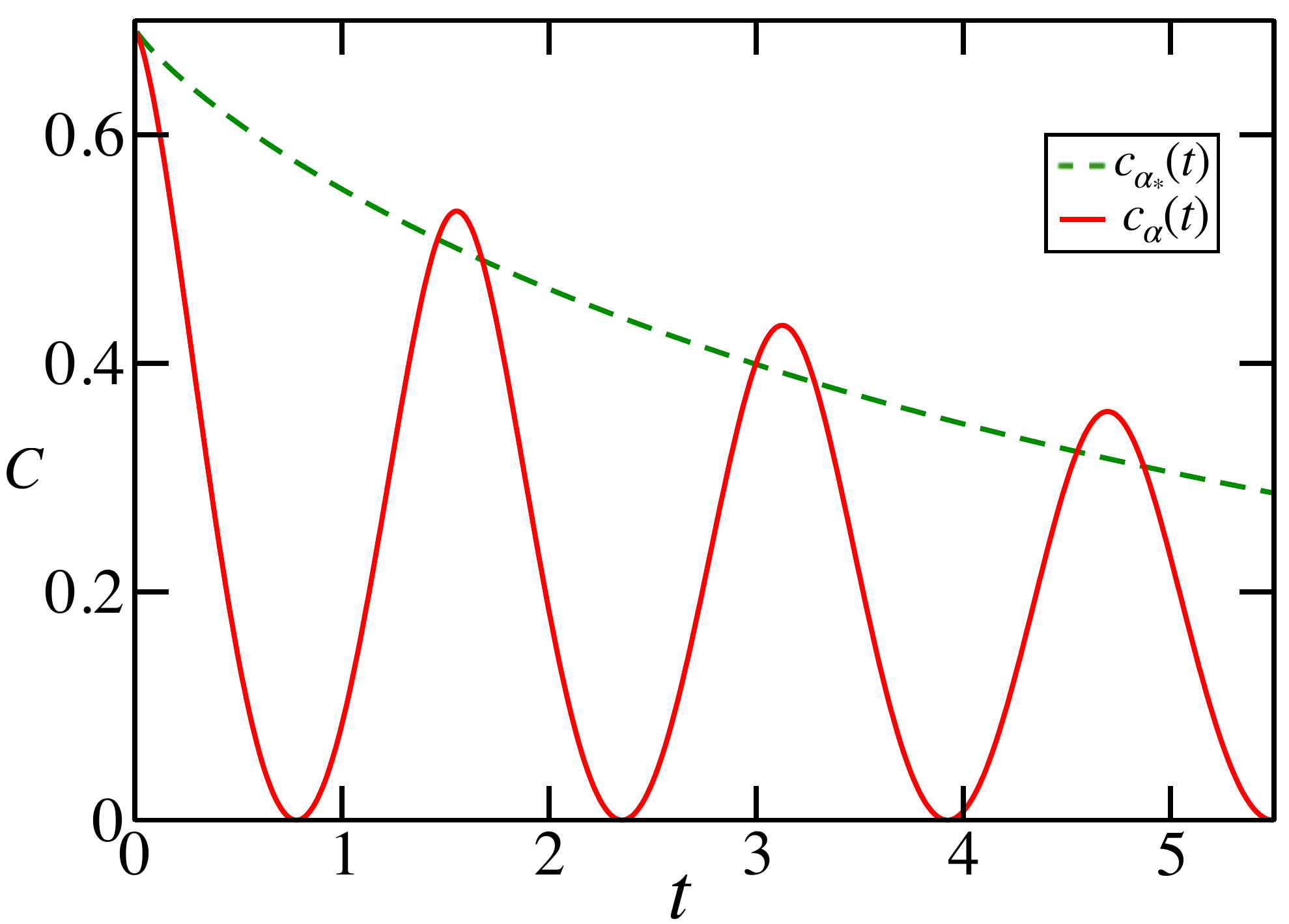}
\caption{
Illustration of an increase in classical capacity for the qubit evolution with  $r=1/(3\mathrm{s})$, and $\omega=2/\mathrm{s}$.
}
\label{kernel4}
\end{figure}

As a case study, we analyze the evolution where the Markovian part is given by the generator
\begin{equation}
\mathcal{L}^{\rm M}(t)=
\frac{r}{d+e^{rt}}\sum_{\alpha=1}^{d+1}\mathcal{L}_\alpha
\end{equation}
with $r>0$. The solution reads
\begin{equation}
\lambda^{\rm M}_\alpha(t)=\frac{1+de^{-rt}}{d+1},
\end{equation}
and $\Lambda^{\rm M}(t)$ is always completely positive. One finds that the corresponding kernel has the eigenvalues
\begin{equation}
\kappa^{\rm M}_\alpha(t)=-\frac{dr}{d+1}\left(\delta(t)-\frac{r}{d+1}
e^{-\frac{rt}{d+1}}\right).
\end{equation}
Therefore, from the kernel point of view, our generalization means that the Markovian part of the kernel has not only terms proportional to the Dirac delta but also some purely non-local parts. The environmental noise is realized with $\kappa_{\alpha}^{\rm N}(t)$ from Eq. (\ref{ker}) for a fixed $Z=\frac{r}{d+1}$. The associated solution is $\lambda^{\rm N}_{\alpha_\ast}(t)=1$ and
\begin{equation}\label{L7}
\lambda^{\rm N}_{\alpha}(t)=\frac{2\omega}{P}e^{-\frac{rt}{2(d+1)}}
\cos\left(\frac{Pt}{2}-\arctan\frac{r}{P(d+1)}\right)
\end{equation}
for $\alpha\neq\alpha_\ast$, where $P=\sqrt{4\omega^2-r^2/(d+1)^2}$. For the complete positivity condition, see Eq. (\ref{CPC}). Finally, the dynamical map generated by $K(t)=\mathfrak{K}(t)+\mathbb{K}(t)$ is characterized by $\lambda_{\alpha_\ast}(t)=\lambda^{\rm M}_\alpha(t)$ and
\begin{equation}
\lambda_\alpha(t)=\frac{2X}{(d+1)Y}e^{-\frac{rt}{2}}\cos\left(\frac{Yt}{2}
+\arctan\frac{r(d-1)}{Y(d+1)}\right),
\end{equation}
where $\alpha\neq\alpha_\ast$, $Y=\sqrt{4\omega^2-r^2}$, and $X=\sqrt{(d+1)^2\omega^2-dr^2}$. For this map to describe a physical evolution, it is enough that
\begin{equation}
\frac{X}{Y}\leq\frac{1}{d-1}e^{rt/2}+\frac{1}{2}e^{-rt/2}
\end{equation}
with the first minimum of the cosine function corresponding to
\begin{equation}
t_\ast=\frac{2}{Y}\left(\pi-\arctan\frac{(d-1)r}{(d+1)Y}\right).
\end{equation}
Analogically to the previous example, the lower bound for the classical capacity of $\Lambda(t)$ is given by
\begin{equation}
C[\Lambda(t)]=\max\Big\{c_\alpha(t),c_{\alpha_\ast}(t)\Big\},
\end{equation}
for $c_\alpha(t)$ defined in Eq. (\ref{33}), where $C[\Lambda^{\rm M}(t)]=c_{\alpha_\ast}(t)$ is the capacity of the Markovian evolution.
Again, we observe a temporary increase in the channel capacity for a certain set of kernel parameters (see Fig. \ref{kernel4} for the qubit evolution).

\FloatBarrier

\section{Conclusions}

We analyzed the classical capacity of the generalized Pauli channels generated via the memory kernel master equations. We compared the evolution of channel capacity for the Markovian semigroup and for the dynamical map generated via the memory kernel that is a sum of the Markovian part and the noise part. Note that the local part is legitimate and identical for both maps. The non-local part, which corresponds to environmental noise, was chosen in such a way that the dynamical map that solves the associated Nakajima-Zwanzig equation describes a valid physical evolution. It turns out that the introduction of noise into the master equation can lead to a temporary increase of the classical capacity. In other words, the noise effects can be beneficial in quantum information processing, as they result in an enhanced ability of a quantum channel to reliably transmit classical information. Similar results are obtained after a generalization of the Markovian semigroup to a Markovian evolution provided by a time-local generator. However, we showed that analogical observations cannot be made for time-local master equations. A dynamical map generated via a sum of two time-local generators never produces the classical capacity that is higher than that of a single generator.

It would be interesting to further analyze this topic by considering the kernels for noninvertible Markovian dynamical maps mixed with the noise kernels. Another open question concerns the relation between quantum maps that increase classical capacity and maps that increase the channel fidelity. One could expect that capacity enhancement means higher fidelity but not the other way around. A comparative analysis could also be performed for other important measures, like output purity, concurrence, logarithmic negativity, or von Neumann entropy.

\section{Acknowledgements}

This paper was supported by the Polish National Science Centre project No. 2018/30/A/ST2/00837. Additionally, K.S. was supported by the Foundation for Polish Science (FNP).

\bibliography{C:/Users/cynda/OneDrive/Fizyka/bibliography}
\bibliographystyle{C:/Users/cynda/OneDrive/Fizyka/beztytulow2}

%

%
%
%

\end{document}